\documentclass[aps,prb,twocolumn,showpacs]{revtex4}

\usepackage{graphicx}
\usepackage{amsmath}
\usepackage{amssymb}

\begin{document}

\title{Electronic structure of cuprate superconductors in a full charge-spin recombination scheme}

%\title{Electronic structure of cuprate superconductors}

\author{Shiping Feng$^{*}$ and L\"ulin Kuang}

\affiliation{Department of Physics, Beijing Normal University, Beijing 100875, China}

\author{Huaisong Zhao}

\affiliation{College of Physics, Qingdao University, Qingdao 266071, China}

\begin{abstract}
A long-standing unsolved problem is how a microscopic theory of superconductivity in cuprate superconductors based on the charge-spin separation can produce a large electron Fermi surface. Within the framework of the kinetic-energy driven superconducting mechanism, a full charge-spin recombination scheme is developed to fully recombine a charge carrier and a localized spin into a electron, and then is employed to study the electronic structure of cuprate superconductors in the superconducting-state. In particular, it is shown that the underlying electron Fermi surface fulfills Luttinger's theorem, while the superconducting coherence of the low-energy quasiparticle excitations is qualitatively described by the standard d-wave  Bardeen-Cooper-Schrieffer formalism. The theory also shows that the observed peak-dip-hump structure in the electron spectrum and Fermi arc behavior in the underdoped regime are mainly caused by the strong energy and momentum dependence of the electron self-energy.
\end{abstract}


\maketitle

\section{Introduction}

Superconductivity in cuprate superconductors occurs upon charge carrier doping Mott insulators \cite{Bednorz86}. Experimentally, it is well established that the Mott insulating state at zero doping develops long-range order antiferromagnetism \cite{Fujita12,Dean14}. Upon doping the antiferromagnetic (AF) long-range order (AFLRO) disappears rapidly and soon thereafter is replaced by the superconducting (SC) ground-state \cite{Fujita12,Dean14}. After intensive investigations over more than two decades, a large body of data available from a wide variety of measurement techniques have introduced important constraints on the microscopic model and SC theory \cite{Fujita12,Dean14,Damascelli03,Campuzano04,Kordyuk10,Hufner08,Kastner98,Timusk99,Basov05}. At the temperature above the SC transition-temperature $T_{\rm c}$, the electron is in a normal-state, however, the normal-state in the underdoped and optimally doped regimes is not normal at all, since the normal-state of cuprate superconductors exhibits a number of the anomalous properties \cite{Hufner08,Kastner98,Timusk99,Basov05} in the sense that they do not fit in with the standard Landau-Fermi liquid theory. The defining characteristic is that the resistivity grows nearly linearly with temperature \cite{Kastner98,Timusk99,Basov05,Batlogg94}. Infrared measurements confirmed that the resistivity scales similarly with energy and temperature, and that the anomalous energy dependence extends up to an energy equivalent to at least 300 meV \cite{Kastner98,Timusk99,Basov05,Batlogg94}. More importantly, the conductivity in the underdoped and optimally doped regimes shows a non-Drude behavior (the conductivity decays as $\rightarrow 1/\omega$) at low energies  \cite{Timusk99,Basov05}, and is carried by $\delta$ charge carriers, where $\delta$ is the charge carrier doping concentration. These are strong experimental evidences supporting the notion of the charge-spin separation \cite{Anderson00,Lee99}. Superconductivity is an instability of the normal-state. However, one of the most striking dilemmas is that the SC coherence of the low-energy quasiparticle excitations in cuprate superconductors seems to be described by the standard Bardeen-Cooper-Schrieffer (BCS) formalism \cite{Campuzano96,Matsui03,Balatsky07,Fujita08,Balatsky09}, although the normal-state is undoubtedly not the standard Landau Fermi-liquid on which the conventional electron-phonon theory is based. Angle-resolved photoemission spectroscopy (ARPES) experiments reveal sharp spectral peaks in the single-particle excitation spectrum \cite{Damascelli03,Campuzano04,Campuzano96,Matsui03,Balatsky07,Fujita08,Balatsky09,Campuzano99,Fedorov99,Ding01,DLFeng00}, indicating the presence of quasiparticle-like states, which is also consistent with the long lifetime of the electronic state as it has been determined by the conductivity measurements \cite{Timusk99,Basov05}. In particular, as a direct method for probing the electron Fermi surface, the early ARPES measurements indicate that in the entire doping range, the underlying electron Fermi surface satisfies Luttinger's theorem \cite{Takahashi89,Campuzano90,Olson90,Ding97}, i.e., the electron Fermi surface with the area is proportional to $1-\delta$. Later, the ARPES experimental studies show that in the underdoped and optimally doped regimes, although the antinodal region of the electron Fermi surface is gapped out, leading to the notion that only part of the electron Fermi surface survives as the disconnected Fermi arcs around the nodes \cite{Norman98,Kanigel06,Nakayama09,Meng11,Yoshida06}, the underlying electron Fermi surface determined from the low-energy spectral weight still fulfills Luttinger's theorem in the entire doping range \cite{Yoshida06}. These ARPES experimental facts \cite{Campuzano96,Matsui03,Balatsky07,Fujita08,Balatsky09,Campuzano99,Fedorov99,Ding01,DLFeng00,Takahashi89,Campuzano90,Olson90,Ding97,Norman98,Kanigel06,Nakayama09,Meng11,Yoshida06} on the other hand provide strong evidences supporting the notion of the charge-spin recombination \cite{Anderson00,Lee99}. Since the electron Fermi surface is a fundamental property of interacting electron systems, the study of the nature of the electron Fermi surface should be crucial for understanding the electronic structure of cuprate superconductors.

Theoretically, it is generally agreed that the strong electron correlation plays a dominant role in the description of the anomalous normal-state properties of cuprate superconductors and the related SC mechanism \cite{Anderson87,Phillips10}. This strong electron correlation originates from a large on-site repulsion between two electrons occupying the same site, which effectively translates into an elimination of double electron occupancy. Apart from the numerical techniques, an intuitively appealing approach to implement this elimination of double electron occupancy and the charge-spin separation is the slave-particle approach \cite{Lee99,Kotliar88,Yu92}, where the constrained electron operator $C_{l\sigma}$ is given by a composite operator as $C_{l\sigma}=a^{\dagger}_{l}f_{l\sigma}$, with $a^{\dagger}_{l}$ as the slave boson and $f_{l\sigma}$ as the fermion or {\it vice versa}, i.e., $a^{\dagger}_{l}$ as the fermion and $f_{l\sigma}$ as the boson. In this slave-particle approach, the operator $f_{l\sigma}$ carries spin index (spinon) and the operator $a^{\dagger}_{l}$ is interpreted as creating a vacancy (holon). The elimination of double electron occupancy is presented by the requirement that $a^{\dagger}_{l}a_{l}+ \sum_{\sigma} f^{\dagger}_{l\sigma} f_{l\sigma}=1$ which can be enforced by introducing a Lagrangian multiplier. Moreover, the doped holes are mainly responsible for the charge transport, and the relaxation time of the excitation from the spin degree of freedom is mainly responsible for the spin response, while the SC-state is characterized by the charge-spin recombination, forming SC quasiparticles \cite{Lee99,Yu92}. In the conventional charge-spin recombination scheme, the electron Green's function in space-time is a product of the holon and spinon Green's functions \cite{Anderson00,Lee99,Yu92,Feng93}. The resulting Fourier transform is a convolution of the holon and spinon Green's functions. However, in the early days of superconductivity, we \cite{Feng93} have formally proved that the electron Fermi surface observed experimentally from the ARPES experiments can not be restored based on the conventional charge-spin recombination. In this case, a long-standing unsolved problem is how a microscopic theory based on the charge-spin separation can give a consistent description of the electronic structure of cuprate superconductors in terms of a full charge-spin recombination. By the full charge-spin recombination we refer to the obtained electron propagator that can produce a large electron Fermi surface with an area proportional to $1-\delta$. For a proper treatment of the strong electron correlation in cuprate superconductors, we \cite{Feng9404,Feng15} have developed a fermion-spin theory based on the charge-spin separation, where the constrained electron operator is decoupled as a product of a charge carrier and a localized spin, and then the electron motion is restricted in the restricted Hilbert space without double electron occupancy. Within the framework of the fermion-spin theory, we \cite{Feng15,Feng0306,Feng12} have established a kinetic-energy driven SC mechanism, where the charge-carrier pairing state is conventional BCS-like with the d-wave symmetry, although the pairing mechanism is driven by the kinetic-energy by the exchange of spin excitations in the higher powers of the doping concentration. In particular, this kinetic-energy driven charge-carrier pairing state is controlled by both the charge-carrier pair gap and charge-carrier quasiparticle coherence, which leads to that the charge-carrier pair transition-temperature $T_{\rm c}$ takes a domelike shape with the underdoped and overdoped regimes on each side of the optimal doping, where $T_{\rm c}$ reaches its maximum. Following this kinetic-energy driven SC mechanism, we in this paper develop a full charge-spin recombination scheme, and then show that although the electron Cooper pairing state (then the SC-state) originates from the charge-carrier pairing state, the low-energy excitation of cuprate superconductors in the SC-state resembles the BCS-Bogoliubov quasiparticle. In particular, we show that the obtained formalism for the electron pairing can be used to compute the electronic structure of cuprate superconductors on the first-principles basis much as can be done for conventional superconductors. Moreover, the theory produces a large electron Fermi surface with the area that is given by $1-\delta$, while the striking feature of the peak-dip-hump structure in the single-particle excitation spectrum around the antinodal point and remarkable Fermi arc behavior in the underdoped and optimally doped regimes are mainly caused by the strong energy and momentum dependence of the electron self-energy.

The rest of this paper is organized as follows. Since the work in this paper builds on the kinetic-energy driven SC mechanism, a short summary of the formalism of the kinetic-energy driven SC mechanism is first given in Section \ref{framework}, and then based on this kinetic-energy driven SC mechanism, the basic formalism of the full charge-spin recombination is presented, which is manifested itself by the self-consistent equations that are satisfied by the full electron diagonal and off-diagonal Green's functions. Moreover, we confirm that the SC transition-temperature $T_{\rm c}$ obtained from this full charge-spin recombination scheme is identical to the charge-carrier pair transition-temperature. In Section \ref{electronic-structure}, the full electron diagonal Green's function is employed to study the electronic structure of cuprate superconductors in the SC-state, and then some main features of the ARPES measurements on cuprate superconductors in the SC-state are qualitatively reproduced. Finally, we give a summary in Section \ref{conclusions}.

\section{Theoretical framework}\label{framework}

Superconductivity in cuprate superconductors is found in copper oxide-based compounds with a layered crystal structure consisting of the two-dimensional CuO$_{2}$ planes separated by insulating layers \cite{Bednorz86,Fujita12,Dean14,Damascelli03,Campuzano04}. The key element shared by all such structure is the CuO$_{2}$, and then it seems evident that the relatively high $T_{\rm c}$ in cuprate superconductors is dominated by this CuO$_{2}$ plane \cite{Bednorz86,Fujita12,Dean14,Damascelli03,Campuzano04}. Immediately following the discovery of superconductivity in cuprate superconductors, Anderson \cite{Anderson87} argued that the essential physics of the doped CuO$_{2}$ plane is contained in the $t$-$J$ model on a square lattice,
\begin{eqnarray}\label{tJmodel}
H&=&-t\sum_{l\hat{\eta}\sigma}C^{\dagger}_{l\sigma}C_{l+\hat{\eta}\sigma}+t'\sum_{l\hat{\tau}\sigma}C^{\dagger}_{l\sigma}C_{l+\hat{\tau}\sigma}+\mu\sum_{l\sigma}
C^{\dagger}_{l\sigma}C_{l\sigma}\nonumber\\
&+&J\sum_{l\hat{\eta}}{\bf S}_{l}\cdot {\bf S}_{l+\hat{\eta}},
\end{eqnarray}
where the summation is over all sites $l$, and for each $l$, over its nearest-neighbors $\hat{\eta}$ or the next nearest-neighbors $\hat{\tau}$, $C^{\dagger}_{l\sigma}$ ($C_{l\sigma}$) is electron creation (annihilation) operator with spin $\sigma$, ${\bf S}_{l}=(S^{\rm x}_{l},S^{\rm y}_{l},S^{\rm z}_{l})$ are spin operators, and $\mu$ is the chemical potential. In spite of its simple form, the $t$-$J$ model (\ref{tJmodel}) is proved to be very difficult to analyze, analytically as well as numerically. However, the most difficult in the analytical treatment of the $t$-$J$ model (\ref{tJmodel}) comes mainly from the local constraint of no double electron occupancy, i.e., $\sum_{\sigma} C^{\dagger}_{l\sigma}C_{l\sigma}\leq 1$, while the strong electron correlation manifests itself by this local constraint of no double electron occupancy, and therefore the crucial requirement is to impose this local constraint. In order to satisfy this local constraint, we employ the fermion-spin formalism \cite{Feng9404,Feng15}, in which the electron operators $C_{l\uparrow}$ and $C_{l\downarrow}$ are replaced by,
\begin{eqnarray}\label{CSS}
C_{l\uparrow}=h^{\dagger}_{l\uparrow}S^{-}_{l}, ~~~~ C_{l\downarrow}=h^{\dagger}_{l\downarrow}S^{+}_{l},
\end{eqnarray}
respectively. The spinful fermion operator $h_{l\sigma}=e^{-i\Phi_{l\sigma}}h_{l}$ keeps track of the charge degree of freedom of the constrained electron together with some effects of spin configuration rearrangements due to the presence of the doped hole itself (charge carrier), while the spin operator $S_{l}$ represents the spin degree of freedom of the constrained electron, and then the local constraint of no double occupancy is satisfied at each site. In this fermion-spin representation (\ref{CSS}), the original $t$-$J$ model (\ref{tJmodel}) can be expressed explicitly as,
\begin{eqnarray}\label{CSStJmodel}
H&=&t\sum_{l\hat{\eta}}(h^{\dagger}_{l+\hat{\eta}\uparrow}h_{l\uparrow}S^{+}_{l}S^{-}_{l+\hat{\eta}}+h^{\dagger}_{l+\hat{\eta}\downarrow}h_{l\downarrow}S^{-}_{l}S^{+}_{l+\hat{\eta}}) \nonumber\\
&-&t'\sum_{l\hat{\tau}}(h^{\dagger}_{l+\hat{\tau}\uparrow}h_{l\uparrow}S^{+}_{l}S^{-}_{l+\hat{\tau}}+h^{\dagger}_{l+\hat{\tau}\downarrow}h_{l\downarrow}S^{-}_{l}S^{+}_{l+\hat{\tau}})
\nonumber\\
&-&\mu\sum_{l\sigma} h^{\dagger}_{l\sigma}h_{l\sigma}+J_{{\rm eff}}\sum_{l\hat{\eta}}{\bf S}_{l}\cdot {\bf S}_{l+\hat{\eta}},
\end{eqnarray}
where $J_{{\rm eff}}=(1-\delta)^{2}J$, and $\delta=\langle h^{\dagger}_{l\sigma}h_{l\sigma}\rangle=\langle h^{\dagger}_{l}h_{l}\rangle$ is the charge-carrier doping concentration. Since the $t$-$J$ model (\ref{tJmodel}) is obtained from the large-$U$ Hubbard model, taking the large-$U$ limit and making certain approximations \cite{Gros87}, there is a mixing of kinetic energy and potential energy (Coulombic contribution) \cite{Singh07} in going from the large-$U$ Hubbard model to the $t$-$J$ model (\ref{tJmodel}), i.e., the original kinetic energy in the large-$U$ Hubbard model has been reorganized as the kinetic energy of the lower Hubbard band in the $t$-$J$ model (\ref{tJmodel}), which therefore contains a strong Coulombic contribution due to the restriction of no double occupancy of a given site. However, as a consequence of the the $t$-$J$ model (\ref{CSStJmodel}) in the fermion-spin representation, the mixing of kinetic energy and potential energy in the kinetic-energy term of the original $t$-$J$ model (\ref{tJmodel}) has been released as the interaction between charge carriers and spins in the $t$-$J$ model (\ref{CSStJmodel}), which therefore dominates the essential physics of cuprate superconductors, while the magnetic exchange-energy term is to form an adequate spin configuration only. In particular, this spin configuration is strongly rearranged due to the effect of the charge-carrier hopping $t$ on the spins. In the $t$-$J$ model (\ref{CSStJmodel}), the charge-carrier quasiparticle and spin excitation are strongly renormalized each other because of the coupling between the two degrees of freedom. In this case, three basic low-energy excitations for the charge-carrier quasiparticle, spin excitation, and electron quasiparticle, respectively, emerge as the propagating modes in a doped Mott insulator \cite{Phillips10,Feng15}, with the scattering of charge-carrier quasiparticles due to spin fluctuations that mainly governs the charge transport, and the scattering of spin excitations due to charge-carrier fluctuations dominates the spin response, while as a natural result of the charge-spin recombination, the electron quasiparticles are responsible for the electronic structure.

\subsection{Kinetic-energy driven superconducting mechanism}\label{Brief-review}

It is commonly believed that the existence of charge-carrier pairs is the hallmark of superconductivity \cite{Cooper56,Anderson07}, since these charge-carrier pairs behave as effective bosons, and can form something analogous to a Bose condensate that flows without resistance. However, the pairing means that there is an attraction between charge carriers. In conventional superconductors, the effective attraction between charge carriers (electrons) generates by the exchange of phonons, which act like a bosonic glue to hold the charge-carrier pairs together \cite{Bardeen57}, and then these charge-carrier pairs condense into the SC-state. In this conventional electron-phonon SC mechanism \cite{Bardeen57}, the wave function for the pairs turns out to be peaked at zero separation of the charge carriers, and then the SC-state has a s-wave symmetry. On the other hand, the SC-state in cuprate superconductors is also characterized by the charge-carrier pairs \cite{Tsuei00}, with an energy gap in the single-particle excitation spectrum. However, as a natural consequence of the unconventional SC mechanism that is responsible for the high $T_{\rm c}$, the charge-carrier pair in cuprate superconductors has a dominant d-wave symmetry \cite{Tsuei00}. This d-wave SC-state also implies that there is a strongly momentum-dependent attraction between charge carriers without phonons \cite{Monthoux07}. In particular, a large body of experimental data \cite{Fujita12,Dean14,Damascelli03,Campuzano04,Kordyuk10,Hufner08,Kastner98,Timusk99,Basov05} available from a wide variety of measurement techniques benefits that the spin excitation, which is a generic consequence of strong electron correlation, can mediate the charge-carrier pairing state in cuprate superconductors in analogy to the phonon-mediate pairing mechanism in conventional superconductors. Within the framework of the fermion-spin theory (\ref{CSS}), we \cite{Feng15,Feng0306,Feng12} have established a kinetic-energy driven SC mechanism, where the attractive interaction between charge carriers originates directly from the interaction between charge carriers and spins in the kinetic energy of the $t$-$J$ model (\ref{CSStJmodel}) by the exchange of spin excitations in the higher powers of the doping concentration. This attractive interaction leads to the formation of the charge-carrier pairs, while the electron Cooper pairs originate from the charge-carrier pairing state are due to the charge-spin recombination, and then these electron Cooper pairs condense into the d-wave SC-state. Our work of the full charge-spin recombination and its application to study the electronic structure of cuprate superconductors in the SC-state builds on the kinetic-energy driven SC mechanism in Refs. \cite{Feng15,Feng0306,Feng12}, and only a short summary of the formalism is therefore given in this subsection for convenience in the following discussions. In our previous discussions in the doped regime without AFLRO, the full charge-carrier diagonal and off-diagonal Green's functions of the $t$-$J$ model in the charge-carrier pairing state satisfy following self-consistent equations \cite{Feng15,Feng0306,Feng12},
\begin{subequations}\label{HGF}
\begin{eqnarray}
g({\bf k},\omega)&=&g^{(0)}({\bf k},\omega)+g^{(0)}({\bf k},\omega)[\Sigma^{({\rm h})}_{1}({\bf k},\omega)g({\bf k},\omega)\nonumber\\
&-&\Sigma^{({\rm h})}_{2}({\bf k},-\omega) \Gamma^{\dagger}({\bf k},\omega)],~~~~~~~ \label{HDGF}\\
\Gamma^{\dagger}({\bf k},\omega)&=&g^{(0)}({\bf k},-\omega)[\Sigma^{({\rm h})}_{1}({\bf k},-\omega)\Gamma^{\dagger}({\bf k},-\omega)\nonumber\\
&+&\Sigma^{({\rm h})}_{2}({\bf k},-\omega)g({\bf k}, \omega)],~~~~~~~\label{HODGF}
\end{eqnarray}
\end{subequations}
where the mean-field (MF) charge-carrier Green's function $g^{(0)-1}({\bf k},\omega)=\omega-\xi_{\bf k}$, with the charge-carrier excitation spectrum $\xi_{\bf k}$ that has been given in Refs. \cite{Feng9404,Feng15,Feng0306,Feng12}, while the charge-carrier self-energies $\Sigma^{({\rm h})}_{1}({\bf k},\omega)$ in the particle-hole channel and $\Sigma^{({\rm h})}_{2}({\bf k},\omega)$ in the particle-particle channel have been evaluated from the spin bubble as,
%\begin{widetext}
\begin{subequations}\label{HSE}
\begin{eqnarray}
\Sigma^{({\rm h})}_{1}({\bf k},i\omega_{n})&=&{1\over N^{2}}\sum_{{\bf p,p'}}\Lambda^{2}_{{\bf p}+{\bf p}'+{\bf k}}{1\over \beta}\sum_{ip_{m}}g({\bf p}+{\bf k},ip_{m}+i\omega_{n}) \nonumber\\
&\times& \Pi({\bf p},{\bf p}',ip_{m}), \label{H-self-energy-PH}\\
\Sigma^{({\rm h})}_{2}({\bf k},i\omega_{n})&=&{1\over N^{2}}\sum_{{\bf p,p'}}\Lambda^{2}_{{\bf p}+{\bf p}'+{\bf k}}{1\over \beta}\sum_{ip_{m}}\Gamma^{\dagger}({\bf p}+{\bf k}, ip_{m}+i\omega_{n}) \nonumber\\
&\times& \Pi({\bf p},{\bf p}',ip_{m}), \label{H-self-energy-PP}
\end{eqnarray}
\end{subequations}
%\end{widetext}
respectively, with $\Lambda_{{\bf k}}=Zt\gamma_{\bf k}-Zt'\gamma_{\bf k}'$, $\gamma_{{\bf k}}=({\rm cos}k_{x}+{\rm cos}k_{y})/2$, and the spin bubble,
\begin{eqnarray}\label{SB}
\Pi({\bf p},{\bf p}',ip_{m})&=&{1\over\beta}\sum_{ip'_{m}}D^{(0)}({\bf p'},ip_{m}')\nonumber\\
&\times& D^{(0)}({\bf p}'+{\bf p},ip_{m}'+ip_{m}),
\end{eqnarray}
where the MF spin Green's function has been obtained explicitly as,
\begin{eqnarray}
D^{(0)}({\bf k},\omega)&=&{B_{\bf k}\over 2\omega_{\bf k}}\left ({1\over \omega-\omega_{\bf k}}-{1\over\omega+\omega_{\bf k}}\right ),\label{MFSGF}
\end{eqnarray}
with the MF spin excitation spectrum $\omega_{\bf k}$ and the function $B_{\bf k}$ that have been given explicitly in Ref. \cite{Feng9404}. In particular, the charge-carrier self-energy $\Sigma^{({\rm h})}_{2}({\bf k},\omega)$ is identified as the charge-carrier pair gap, while the charge-carrier quasiparticle coherent weight $Z_{\rm hF}$ is closely related to the antisymmetric part of the charge-carrier self-energy $\Sigma^{({\rm h})}_{1}({\bf k},\omega)$. In the case of only discussions of the low-energy sector, the charge-carrier pair gap and charge-carrier quasiparticle coherent weight are obtained in the static-limit approximation as,
\begin{subequations}\label{CCSCE}
\begin{eqnarray}
\bar{\Delta}_{\rm h}({\bf k})&=&\Sigma^{({\rm h})}_{2}({\bf k},\omega)\mid_{\omega=0}=\bar{\Delta}_{\rm h}\gamma^{(\rm d)}_{{\bf k}}, \label{CCSCE-gap} \\
{1\over Z_{\rm hF}}&=&1-{\rm Re}\Sigma^{({\rm h})}_{\rm 1o}({\bf k},\omega=0)\mid_{{\bf k}=[\pi,0]}, \label{CCSCE-WT}
\end{eqnarray}
\end{subequations}
where $\gamma^{(\rm d)}_{{\bf k}}=({\rm cos} k_{x}-{\rm cos}k_{y})/2$. It has been shown that the charge-carrier quasiparticle coherence antagonizes superconductivity, and then $T_{\rm c}$ is depressed to low temperatures \cite{Feng0306}. With the above static-limit approximation in Eq. (\ref{CCSCE}), the full charge-carrier diagonal and off-diagonal Green's functions in Eq. (\ref{HGF}) have been evaluated explicitly as,
\begin{subequations}\label{BCSHGF}
\begin{eqnarray}
g({\bf k},\omega)&=&Z_{\rm hF}\left ({U^{2}_{{\rm h}{\bf k}}\over\omega-E_{{\rm h}{\bf k}}}+{V^{2}_{{\rm h}{\bf k}}\over\omega+E_{{\rm h}{\bf k}}}\right ),\label{BCSHDGF}\\
\Gamma^{\dagger}({\bf k},\omega)&=&-Z_{\rm hF}{\bar{\Delta}_{\rm hZ}({\bf k})\over 2E_{{\rm h}{\bf k}}}\left ({1\over \omega-E_{{\rm h}{\bf k}}}-{1\over\omega
+E_{{\rm h}{\bf k}}} \right ), ~~~~~~\label{BCSHODGF}
\end{eqnarray}
\end{subequations}
where $E_{{\rm h}{\bf k}}=\sqrt{\bar{\xi}^{2}_{{\bf k}}+\mid\bar{\Delta}_{\rm hZ}({\bf k})\mid^{2}}$ is the charge-carrier quasiparticle energy spectrum,
$\bar{\xi}_{{\bf k}}=Z_{\rm hF}\xi_{{\bf k}}$ is the renormalized charge-carrier excitation spectrum, and $\bar{\Delta}_{\rm hZ}({\bf k})=Z_{\rm hF}\bar{\Delta}_{\rm h}({\bf k})$ is the renormalized charge-carrier pair gap, while the charge-carrier quasiparticle coherence factors,
\begin{subequations}\label{BCSHCF}
\begin{eqnarray}
U^{2}_{{\rm h}{\bf k}}={1\over 2}\left (1+{\bar{\xi_{{\bf k}}}\over E_{{\rm h}{\bf k}}}\right ),\\
V^{2}_{{\rm h}{\bf k}}={1\over 2}\left (1-{\bar{\xi_{{\bf k}}}\over E_{{\rm h}{\bf k}}}\right ),
\end{eqnarray}
\end{subequations}
with the constraint $U^{2}_{{\rm h}{\bf k}}+V^{2}_{{\rm h}{\bf k}}=1$ for any wave vector ${\bf k}$ (normalization). In spite of the pairing mechanism driven by the kinetic energy by the exchange of spin excitations, the results in Eqs. (\ref{BCSHGF}) and (\ref{BCSHCF}) are the standard BCS expressions for a d-wave charge-carrier pairing state.

Substituting the full charge-carrier diagonal and off-diagonal Green's functions (\ref{BCSHGF}) into Eq. (\ref{HSE}), the self-energies $\Sigma^{({\rm h})}_{1}({\bf k},\omega)$ and $\Sigma^{({\rm h})}_{2}({\bf k},\omega)$ have been evaluated explicitly. In this case, the self-consistent equations (\ref{CCSCE-gap}) and (\ref{CCSCE-WT}) that are satisfied by the charge-carrier pair gap parameter $\bar{\Delta}_{\rm h}$ and charge-carrier quasiparticle coherent weight $Z_{\rm hF}$, respectively, have been solved simultaneously with other self-consistent equations, and then all order parameters and charge-carrier chemical potential have been determined by the self-consistent calculation without using any adjustable parameters \cite{Feng15,Feng0306,Feng12}. Since the charge-carrier pair order is established through an emergence of the charge-carrier quasiparticle, the charge-carrier pairing state is controlled by both the charge-carrier pair gap $\bar{\Delta}_{\rm h}({\bf k})$ and charge-carrier quasiparticle coherence $Z_{\rm hF}$, which is reflected directly from the self-consistent equations (\ref{CCSCE-gap}) and (\ref{CCSCE-WT}). In particular, the self-consistently calculated result \cite{Huang13,Guo07} of $T_{\rm c}$ shows that the maximal $T_{\rm c}$ occurs around the {\it optimal doping}, and then decreases in both the underdoped and the overdoped regimes, in good agreement with the experimental results of cuprate superconductors \cite{Tallon95}. This microscopic SC theory has given a consistent description of the defining characteristic of cuprate superconductors in the SC-state, including the doping dependence of the conductivity spectrum \cite{Qin14} observed in infrared measurements of the reflectance, the doping dependence of the electromagnetic response \cite{Feng10} obtained by using muon-spin-rotation measurement technique, and the doping dependence of the dynamical spin response \cite{Kuang15} observed in inelastic neutron scattering and resonant inelastic X-ray scattering experiments.

\subsection{Charge-spin recombination}\label{CSRS}

In the kinetic energy driven SC mechanism \cite{Feng15,Feng0306,Feng12}, the charge-carrier pairing state induced by the interaction between charge carriers and spins in the kinetic energy of the $t$-$J$ model (\ref{CSStJmodel}) by the exchange of spin excitations also generates the formation of the electron Cooper pairing state. However, for discussions of the electronic structure of cuprate superconductors in the SC-state, we need to calculate the electron diagonal and off-diagonal Green's functions $G(l-l',t-t')=\langle\langle C_{l\sigma}(t); C^{\dagger}_{l'\sigma}(t')\rangle\rangle$ and $\Im^{\dagger}(l-l',t-t') = \langle\langle C^{\dagger}_{l\uparrow}(t); C^{\dagger}_{l' \downarrow}(t')\rangle\rangle$, which are characterized by the charge-spin recombination. In the fermion-spin theory (\ref{CSS}), the constrained electron is decoupled as a product of a charge carrier and a localized spin. However, in contrast with the charge-spin separation, the purpose of the charge-spin recombination is to recover a constrained electron in terms of the recombination of a charge carrier and a localized spin. In the conventional charge-spin recombination scheme \cite{Anderson00,Lee99,Yu92,Feng93}, the electron Green's function in space-time is a product of the charge-carrier and spin Green's functions, and then the resulting Fourier transform is a convolution of the charge-carrier and spin Green's functions. Based on the conventional charge-spin recombination scheme, we \cite{Feng0306,Huang13,Guo07} have evaluated the electron diagonal and off-diagonal Green's functions, and found that two characteristic features of the charge-carrier pairing state in Eq. (\ref{BCSHGF}) are kept in the electron Cooper pairing state: (A) The electron Cooper pairing state still is conventional BCS-like with the d-wave symmetry; (B) $T_{\rm c}$ is identical to the charge-carrier pair transition-temperature. Within the framework of the kinetic energy driven SC mechanism, the electron Cooper pairing state originates from the charge-carrier pairing state, therefore these two characteristic features of the charge-carrier pairing state also are common properties of the electron Cooper pairing state, and they must be preserved in the electron Cooper pairing state obtained in terms of the charge-spin recombination. However, the conventional charge-spin recombination \cite{Anderson00,Feng93} is a partial charge-spin recombination, since the electron Green's function obtained from this conventional charge-spin recombination does not produce a large electron Fermi surface \cite{Feng93,Guo07}, reflecting that the separated charge carrier and localized spin can not be fully recombined into a constrained electron.

Within the framework of the fermion-spin theory (\ref{CSS}), the coupling between the charge-carrier quasiparticle and spin excitation is manifested itself by the self-consistent equations (\ref{HGF}). Experimentally, it has been established unambiguously that in the doped regime, the electron quasiparticle also couples to the spin excitation \cite{Fujita12,Dean14,Eschrig06,Fong95}. On the other hand, the unique aim of the charge-spin recombination is to recombine the charge-carrier and spin Green's functions into the electron Green's function, which therefore implies that the coupling form between the electron quasiparticle and spin excitation should be the same as that between the charge-carrier quasiparticle and spin excitation, in other words, the self-consistent equations satisfied by the full electron diagonal and off-diagonal Green's functions in the SC-state should be the same with these in Eq. (\ref{HGF}) satisfied by the full charge-carrier diagonal and off-diagonal Green's functions. In this case, we can perform a full charge-spin recombination in which the charge-carrier diagonal and off-diagonal Green's functions in Eq. (\ref{HGF}) are replaced by the electron diagonal and off-diagonal Green's functions $G({\bf k}, \omega)$ and $\Im^{\dagger}({\bf k},\omega)$, respectively, and then the self-consistent equations satisfied by the electron diagonal and off-diagonal Green's functions of the $t$-$J$ model in the SC-state can be obtained explicitly as,
\begin{subequations}\label{EGF}
\begin{eqnarray}
G({\bf k},\omega)&=&G^{(0)}({\bf k},\omega)+G^{(0)}({\bf k},\omega)[\Sigma_{1}({\bf k},\omega)G({\bf k},\omega)\nonumber\\
&-& \Sigma_{2}({\bf k},-\omega)\Im^{\dagger}({\bf k},\omega)], \label{EDGF} \\
\Im^{\dagger}({\bf k},\omega)&=&G^{(0)}({\bf k},-\omega)[\Sigma_{1}({\bf k},-\omega)\Im^{\dagger}({\bf k},-\omega)\nonumber\\
&+&\Sigma_{2}({\bf k},-\omega)G({\bf k}, \omega)], \label{EODGF}
\end{eqnarray}
\end{subequations}
with the MF electron Green's function $G^{(0)}({\bf k},\omega)$ that can be obtained directly from the $t$-$J$ model (\ref{tJmodel}) as,
\begin{eqnarray}\label{MFEGF}
G^{(0)}({\bf k},\omega)&=&{1\over \omega-\varepsilon_{\bf k}},
\end{eqnarray}
where the MF electron excitation spectrum $\varepsilon_{\bf k}=-Zt\gamma_{\bf k}+Zt'\gamma_{\bf k}'+\mu$, with $\gamma_{\bf k}=({\rm cos}k_{x}+{\rm cos}k_{y})/2$, $\gamma_{\bf k}'= {\rm cos} k_{x}{\rm cos}k_{y}$, and $Z$ is the number of the nearest-neighbor or next nearest-neighbor sites on a square lattice, while the electron self-energies $\Sigma_{1}({\bf k}, \omega)$ in the particle-hole channel and $\Sigma_{2}({\bf k},\omega)$ in the particle-particle channel can be obtained directly from the charge-carrier self-energies in Eq. (\ref{HSE}) by the replacement of the charge-carrier diagonal and off-diagonal Green's functions with the corresponding electron diagonal and off-diagonal Green's functions as,
\begin{subequations}\label{ESE}
\begin{eqnarray}
\Sigma_{1}({\bf k},i\omega_{n})&=&{1\over N^{2}}\sum_{{\bf p,p'}}\Lambda^{2}_{{\bf p}+{\bf p}'+{\bf k}}{1\over \beta}\sum_{ip_{m}}G({\bf p}+{\bf k},ip_{m}+i\omega_{n})\nonumber\\ &\times& \Pi({\bf p}, {\bf p}',ip_{m}), \label{electron-self-energy-PH}\\
\Sigma_{2}({\bf k},i\omega_{n})&=&{1\over N^{2}}\sum_{{\bf p,p'}}\Lambda^{2}_{{\bf p}+{\bf p}'+{\bf k}}{1\over \beta}\sum_{ip_{m}}\Im^{\dagger}({\bf p}+{\bf k},ip_{m}+i\omega_{n}) \nonumber\\
&\times& \Pi({\bf p},{\bf p}',ip_{m}). \label{electron-self-energy-PP}
\end{eqnarray}
\end{subequations}
In analogy to the case in the charge-carrier pairing state \cite{Feng15,Feng0306,Feng12}, both the pairing force and electron Cooper pair order parameter have been incorporated into the electron self-energy $\Sigma_{2}({\bf k},\omega)$ in the particle-particle channel, and therefore it is called the electron Cooper pair gap in the electron excitation spectrum, $\bar{\Delta}({\bf k},\omega)= \Sigma_{2}({\bf k},\omega)$, which determines both the quasiparticle energy spectrum and the energy of the condensate. On the other hand, the electron self-energy $\Sigma_{1}({\bf k},\omega)$ in the particle-hole channel renormalizes the MF electron spectrum, and therefore it describes the electron quasiparticle coherence. In particular, the electron self-energy $\Sigma_{2}({\bf k},\omega)$ is an even function of $\omega$, while the electron self-energy $\Sigma_{1}({\bf k},\omega)$ is not. It therefore follows common practice to separate the electron self-energy $\Sigma_{1}({\bf k},\omega)$ into the symmetric and antisymmetric parts, i.e., $\Sigma_{1}({\bf k},\omega)=\Sigma_{\rm 1e} ({\bf k},\omega)+\omega \Sigma_{\rm 1o}({\bf k},\omega)$, and then both $\Sigma_{\rm 1e}({\bf k},\omega)$ and $\Sigma_{\rm 1o}({\bf k},\omega)$ are an even function of $\omega$. Moreover, the antisymmetric part $\Sigma_{\rm 1o}({\bf k},\omega)$ of the electron self-energy $\Sigma_{1}({\bf k},\omega)$ is directly associated with the electron quasiparticle coherent weight as, $Z^{-1}_{\rm F}({\bf k},\omega)=1-{\rm Re}\Sigma_{\rm 1o}({\bf k},\omega)$. Since we only focus on the low-energy behavior, the electron Cooper pair gap and electron quasiparticle coherent weight can be generally discussed in the static-limit approximation, i.e., $\bar{\Delta}({\bf k})= \Sigma_{2}({\bf k},\omega) \mid_{\omega=0}= \bar{\Delta}\gamma^{(\rm d)}_{{\bf k}}$, and $Z^{-1}_{\rm F}({\bf k})=1-{\rm Re}\Sigma_{\rm 1o}({\bf k},\omega)\mid_{\omega=0}$. As in conventional superconductors \cite{Mahan81}, the retarded function ${\rm Re}\Sigma_{\rm 1e}({\bf k},\omega)\mid_{\omega=0}$ just renormalizes the chemical potential. Although $Z_{\rm F}({\bf k})$ still is a function of momentum, however, the wave vector ${\bf k}$ in $Z_{\rm F}({\bf k})$ can be chosen as,
\begin{eqnarray}\label{EQCW}
{1\over Z_{\rm F}}=1-{\rm Re}\Sigma_{\rm 1o}({\bf k},\omega=0)\mid_{{\bf k}=[\pi,0]},
\end{eqnarray}
just as it has been done in the ARPES experiments \cite{Ding01,DLFeng00}. With the help of the above static-limit approximation, the full electron diagonal and off-diagonal Green's functions in Eq. (\ref{EGF}) can be evaluated explicitly as,
\begin{subequations}\label{BCSEGF}
\begin{eqnarray}
G({\bf k},\omega)&=&Z_{\rm F}\left ( {U^{2}_{\bf k}\over\omega-E_{\bf k}}+{V^{2}_{\bf k}\over\omega+E_{\bf k}}\right ),\label{BCSEDGF}\\
\Im^{\dagger}({\bf k},\omega)&=&-Z_{\rm F}{\bar{\Delta}({\bf k})\over 2E_{\bf k}}\left ({1\over\omega-E_{\bf k}}-{1\over\omega+E_{\bf k}}\right ),~~~~~\label{BCSEODGF}
\end{eqnarray}
\end{subequations}
where $E_{\bf k}=\sqrt{\bar{\varepsilon}^{2}_{\bf k}+\mid\bar{\Delta}_{\rm Z}({\bf k})\mid^{2}}$ is the electron quasiparticle energy spectrum,
$\bar{\varepsilon}_{\bf k}=Z_{\rm F}\varepsilon_{\bf k}$ is the renormalized electron excitation spectrum, and $\bar{\Delta}_{\rm Z}({\bf k})=Z_{\rm F}\bar{\Delta}({\bf k})$ is the renormalized electron Cooper pair gap, while the electron quasiparticle coherence factors,
\begin{subequations}\label{BCSECF}
\begin{eqnarray}
U^{2}_{\bf k}={1\over 2}\left (1+{\bar{\varepsilon}_{\bf k}\over E_{\bf k}}\right ),\label{BCSCFU}\\
V^{2}_{\bf k}={1\over 2}\left (1-{\bar{\varepsilon}_{\bf k}\over E_{\bf k}}\right ),\label{BCSCFV}
\end{eqnarray}
\end{subequations}
satisfy the sum rule $U^{2}_{\bf k}+V^{2}_{\bf k}=1$ for any wave vector ${\bf k}$ (normalization). In spite of the electron Cooper pairs originated from the charge-carrier pairing state, the standard d-wave BCS formalism in Eqs. (\ref{BCSEGF}) and (\ref{BCSECF}) obtained from the kinetic energy driven SC mechanism have unambiguously indicates the Bogoliubov quasiparticle nature of the low-energy excitations of cuprate superconductors in the SC-state, i.e., the Bogoliubov quasiparticle does not carry definite charge, and is a coherent combination of the electron and its absence, then the SC coherence of the low-energy quasiparticle excitations and the related electronic structure can be discussed on the first-principles basis much as can be done for conventional superconductors. In this case, the characteristic feature (A) of the charge-carrier pairing state is therefore satisfied by the standard d-wave BCS formalism in Eqs. (\ref{BCSEGF}) and (\ref{BCSECF}). In particular, it should be emphasized that the Bogoliubov-type dispersion of cuprate superconductors in the SC-state, and the momentum dependence of the coherence factors $U_{\bf k}$ and $V_{\bf k}$ in Eq. (\ref{BCSECF}) have been confirmed experimentally from the ARPES measurements \cite{Campuzano96,Matsui03,Balatsky07,Fujita08,Balatsky09}. On the other hand, the electron quasiparticle coherent weight $Z_{\rm F}$ reduces the electron quasiparticle bandwidth, and suppresses the spectral weight of the single-particle excitation spectrum, then the energy scale \cite{Guo06} of the electron quasiparticle band is controlled by the magnetic exchange coupling $J$. Moreover, the AF short-range order (AFSRO) correlation has been incorporated into the SC-state through the spin's order parameters entering into the electron self-energies (\ref{ESE}) in the particle-particle and particle-hole channels, therefore there is a coexistence of the SC-state and AFSRO correlation, and then AFSRO fluctuation persists into superconductivity.

Since the full electron diagonal and off-diagonal Green's functions have been obtained explicitly in Eq. (\ref{BCSEGF}), it is straightforward to evaluate the electron self-energies $\Sigma_{1}({\bf k},\omega)$ and $\Sigma_{2}({\bf k},\omega)$ in Eq. (\ref{ESE}) as,
\begin{widetext}
\begin{subequations}\label{ESE1}
\begin{eqnarray}
\Sigma_{1}({\bf k},\omega)&=&{1\over N^{2}}\sum_{{\bf pp'}\nu}(-1)^{\nu+1}\Omega_{\bf pp'k}\left [U^{2}_{{\bf p}+{\bf k}}\left ({F^{(\nu)}_{{\rm 1} {\bf pp'k}} \over\omega+\omega_{\nu{\bf p}{\bf p}'}-E_{{\bf p}+{\bf k}}}+{F^{(\nu)}_{{\rm 2}{\bf pp'k}}\over\omega-\omega_{\nu{\bf p}{\bf p}'}-E_{{\bf p}+{\bf k}}} \right )\right. \nonumber\\
&+&\left . V^{2}_{{\bf p}+{\bf k}}\left ({F^{(\nu)}_{{\rm 1}{\bf pp'k}}\over\omega-\omega_{\nu{\bf p}{\bf p}'}+E_{{\bf p}+{\bf k}}}+{F^{(\nu)}_{{\rm 2}{\bf pp'k}} \over\omega+\omega_{\nu{\bf p}{\bf p}'}+E_{{\bf p}+{\bf k}}}\right )\right ],\label{PHESE}
\end{eqnarray}
\begin{eqnarray}
\Sigma_{2}({\bf k},\omega)&=&{1\over N^{2}}\sum_{{\bf pp'}\nu}(-1)^{\nu}\Omega_{\bf pp'k}{\bar{\Delta}_{\rm Z}({\bf p}+{\bf k})\over 2E_{{\bf p}+  {\bf k}}}\left [\left ({F^{(\nu)}_{{\rm 1}{\bf pp'k}}\over\omega+\omega_{\nu{\bf p}{\bf p}'}-E_{{\bf p}+{\bf k}}}+{F^{(\nu)}_{{\rm 2}{\bf pp'k}}\over\omega-\omega_{\nu{\bf p} {\bf p}'}-E_{{\bf p}+ {\bf k} }} \right )\right .\nonumber\\
&-&\left . \left ({F^{(\nu)}_{{\rm 1}{\bf pp'k}}\over\omega-\omega_{\nu{\bf p}{\bf p}'}+E_{{\bf p}+{\bf k}}}+{F^{(\nu)}_{{\rm 2}{\bf pp'k}}\over\omega+\omega_{\nu{\bf p}{\bf p}'}+ E_{{\bf p}+{\bf k}}}\right )\right ], \label{PPESE}
\end{eqnarray}
\end{subequations}
\end{widetext}
respectively, with $\nu=1,2$, $\Omega_{\bf pp'k}=Z_{\rm F}\Lambda^{2}_{{\bf p}+{\bf p}'+{\bf k}}B_{{\bf p}'}B_{{\bf p}+{\bf p}'}/(4\omega_{{\bf p}'}\omega_{{\bf p}+{\bf p}'})$, $\omega_{\nu{\bf p}{\bf p}'}=\omega_{{\bf p}+{\bf p}'}-(-1)^{\nu}\omega_{\bf p'}$, and the functions,
\begin{subequations}
\begin{eqnarray}
F^{(\nu)}_{{\rm 1}{\bf pp'k}}&=&n_{\rm F}(E_{{\bf p}+{\bf k}})n^{(\nu)}_{{\rm 1B} {\bf pp'}}+n^{(\nu)}_{{\rm 2B}{\bf pp'}}, \\
F^{(\nu)}_{{\rm 2}{\bf pp'k}}&=&[1-n_{\rm F}(E_{{\bf p}+{\bf k}})]n^{(\nu)}_{{\rm 1B}{\bf pp'}}+n^{(\nu)}_{{\rm 2B}{\bf pp'}},
\end{eqnarray}
\end{subequations}
where $n^{(\nu)}_{{\rm 1B}{\bf pp'}}=1+n_{\rm B}(\omega_{{\bf p}'+{\bf p}})+n_{\rm B}[(-1)^{\nu+1}\omega_{\bf p'}]$, $n^{(\nu)}_{{\rm 2B}{\bf pp'}}=n_{\rm B}(\omega_{{\bf p}'+{\bf p}}) n_{\rm B}[(-1)^{\nu+1} \omega_{\bf p'}]$, and $n_{\rm B}(\omega)$ and $n_{\rm F}(\omega)$ are the boson and fermion distribution functions, respectively. In this case, the electron quasiparticle coherent weight $Z_{\rm F}$ and electron Cooper pair gap parameter $\bar{\Delta}$ satisfy following two self-consistent equations,
\begin{widetext}
\begin{subequations}\label{ESCE1}
\begin{eqnarray}
{1\over Z_{\rm F}}&=&1+{1\over N^{2}}\sum_{{\bf pp'}\nu}(-1)^{\nu+1}\Omega_{{\bf pp'}{\bf k}_{\rm A}}\left ({F^{(\nu)}_{{\rm 1}{\bf pp}'{\bf k}_{\rm A}}\over (\omega_{\nu{\bf p}{\bf p}'}-E_{{\bf p}+{\bf k}_{\rm A}})^{2}}+{F^{(\nu)}_{{\rm 2}{\bf pp}'{\bf k}_{\rm A}}\over (\omega_{\nu{\bf p}{\bf p}'}+E_{{\bf p}+{\bf k}_{\rm A}})^{2}}\right ), ~~~~~~~~\label{EQCWSCE}\\
1&=&{4\over N^{3}}\sum_{{\bf pp'k}\nu}(-1)^{\nu}Z_{\rm F}\Omega_{\bf pp'k}{\gamma^{({\rm d})}_{\bf k}\gamma^{({\rm d})}_{{\bf p}+{\bf k}}\over E_{{\bf p}+{\bf k} }}\left ({F^{(\nu)}_{{\rm 1}{\bf pp'k}}\over\omega_{\nu{\bf p}{\bf p}'}-E_{{\bf p}+{\bf k}}}-{F^{(\nu)}_{{\rm 2}{\bf pp'k}}\over \omega_{\nu{\bf p}{\bf p}'}+E_{{\bf p}+{\bf k}}}\right ), ~~~~~~~~\label{EPGPSCE}
\end{eqnarray}
\end{subequations}
\end{widetext}
with ${\bf k}_{\rm A}=[\pi,0]$. These two equations (\ref{EQCWSCE}) and (\ref{EPGPSCE}) must be solved simultaneously with the self-consistent equation,
\begin{eqnarray}\label{ESCE2}
1-\delta &=& {1\over N}\sum_{{\bf k}}Z_{\rm F}\left (1-{\bar{\varepsilon}_{\bf k}\over E_{\bf k}}{\rm tanh}[{1\over 2}\beta E_{\bf k}] \right ),
\end{eqnarray}
then the electron quasiparticle coherent weight $Z_{\rm F}$, the electron Cooper pair gap parameter $\bar{\Delta}$, and the electron chemical potential $\mu$ are determined by the self-consistent calculation without using any adjustable parameters.

\begin{figure}[h!]
\centering
\includegraphics[scale=0.37]{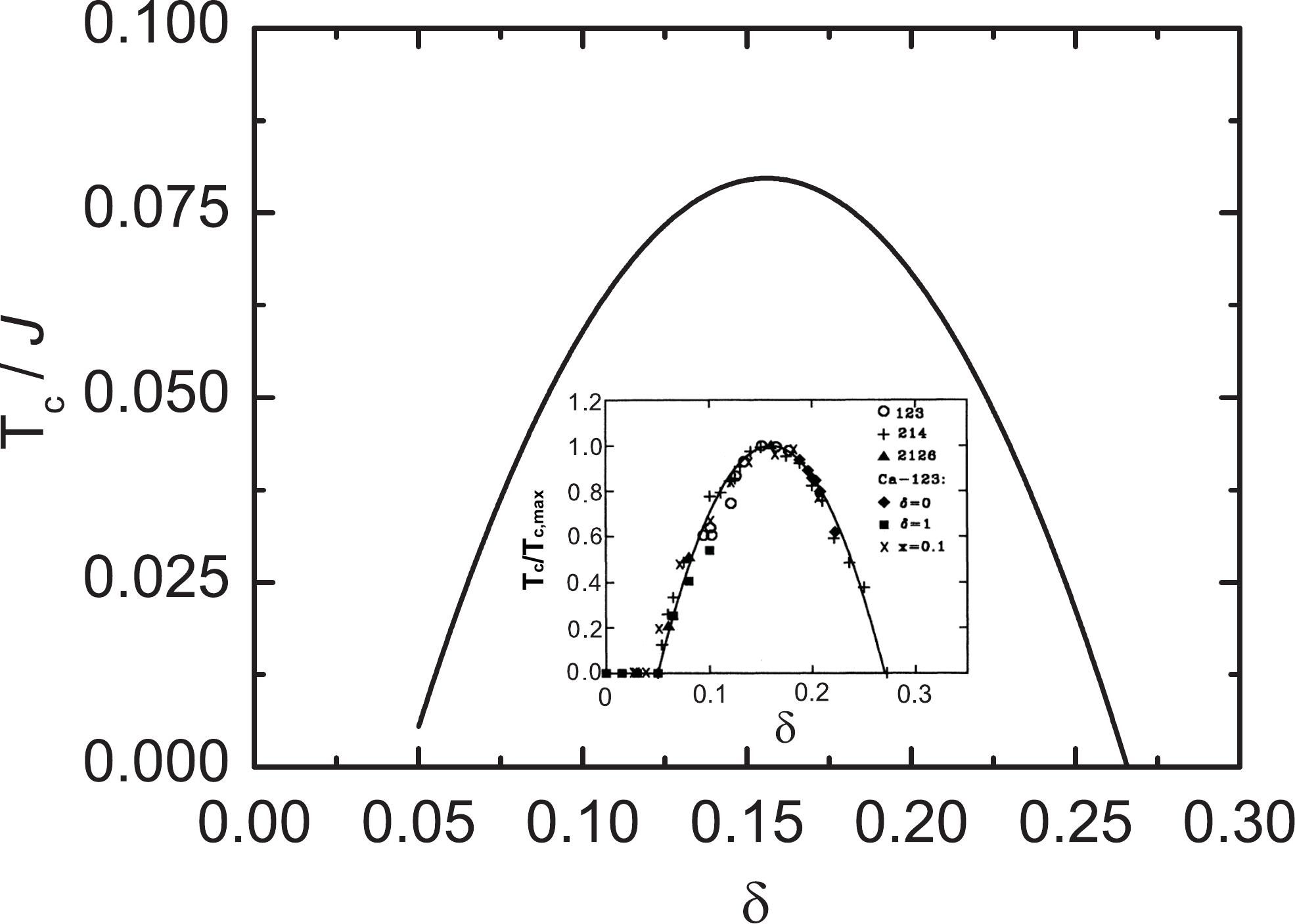}
\caption{$T_{\rm c}$ as a function of doping for $t/J=2.5$ and $t'/t=0.3$. Inset: the corresponding experimental results of cuprate superconductors taken from Ref. [\onlinecite{Tallon95}]. \label{Tc-doping}}
\end{figure}

The above equations (\ref{ESCE1}) and (\ref{ESCE2}) have been calculated self-consistently. In particular, $T_{\rm c}$ has been obtained self-consistently from the self-consistent equations (\ref{ESCE1}) and (\ref{ESCE2}) at the condition $\bar{\Delta}=0$, and the calculated result of $T_{\rm c}$ as a function of doping for $t/J=2.5$ and $t'/t=0.3$ is plotted in Fig. \ref{Tc-doping} in comparison with the corresponding experimental results \cite{Tallon95} (inset). It is shown clearly that $T_{\rm c}$ increases with increasing doping in the underdoped regime, and reaches a maximum around the optimal doping, then decreases with increasing doping in the overdoped regime, in good agreement with the experimental results of cuprate superconductors \cite{Tallon95}. In particular, in a given doping concentration, the value of $T_{\rm c}$ obtained from the electron Cooper pairing state is the same as that obtained from the corresponding charge-carrier pairing state \cite{Huang13}, and therefore the characteristic feature (B) of the charge-carrier pairing state is satisfied exactly. Within the framework of the kinetic-energy driven SC mechanism, the effective attractive interaction between charge carriers originates in their coupling to spin excitations, while in the present full charge-spin recombination scheme in Eqs. (\ref{BCSEGF}) and (\ref{BCSECF}), the electron pairing interaction is mediated by the same spin excitations, which therefore leads to that $T_{\rm c}$ is identical to the charge-carrier pair transition-temperature, and then the domelike shape of the doping dependence of $T_{\rm c}$ with the maximal value appearing around the {\it optimal doping} obtained from the electron Cooper pairing state is a natural consequence of the domelike shape of the doping dependence of the charge-carrier pair transition-temperature with the maximal value appearing around the same {\it optimal doping} obtained from the charge-carrier pairing state \cite{Huang13}. Moreover, in the present full charge-spin recombination scheme, the hole-like charge-carrier quasiparticle coherence factors $V_{{\rm h}{\bf k}}$ and $U_{{\rm h}{\bf k}}$ in Eq. (\ref{BCSHCF}) and the related BCS expressions for the d-wave charge-carrier pairing state in Eq. (\ref{BCSHGF}) have been transferred into the electron quasiparticle coherence factors $U_{\bf k}$ and $V_{\bf k}$ in Eq. (\ref{BCSECF}) and the related d-wave BCS formalism for the electron Cooper pairing state in Eq. (\ref{BCSEGF}), respectively, which means that the d-wave charge-carrier pairs condense in a wide range of the doping concentration, then the electron Cooper pairs originated from the charge-carrier pairing state are due to the full charge-spin recombination, and their condensation automatically gives the electron quasiparticle character. In this case, as in conventional superconductors, the quasiparticles of the electron Cooper pairs in cuprate superconductors also are the excitation of a single electron {\it dressed} with the attractive interaction between paired electrons.

\section{Electronic structure}\label{electronic-structure}

The electronic state in a solid is characterized by its energy dispersion as well as the characteristic lifetime (then the characteristic scattering rate) of an electron placed into such a state. This state is just represented by the electron Green's function, while the electron spectral function is directly related to the analytically continued electron Green's function as $A({\bf k},\omega)=-2{\rm Im}G({\bf k},\omega)$. Since many of the physical properties have often been attributed to particular characteristics of the low-energy quasiparticle excitations determined by the electron spectral function \cite{Damascelli03,Campuzano04,Kordyuk10}, then a central issue to clarify the nature of the physical properties is how the electron spectrum evolves with the doping concentration.

\subsection{Doping dependence of electron spectrum}\label{electron-spectrum}

\begin{figure}[h!]
\centering
\includegraphics[scale=0.35]{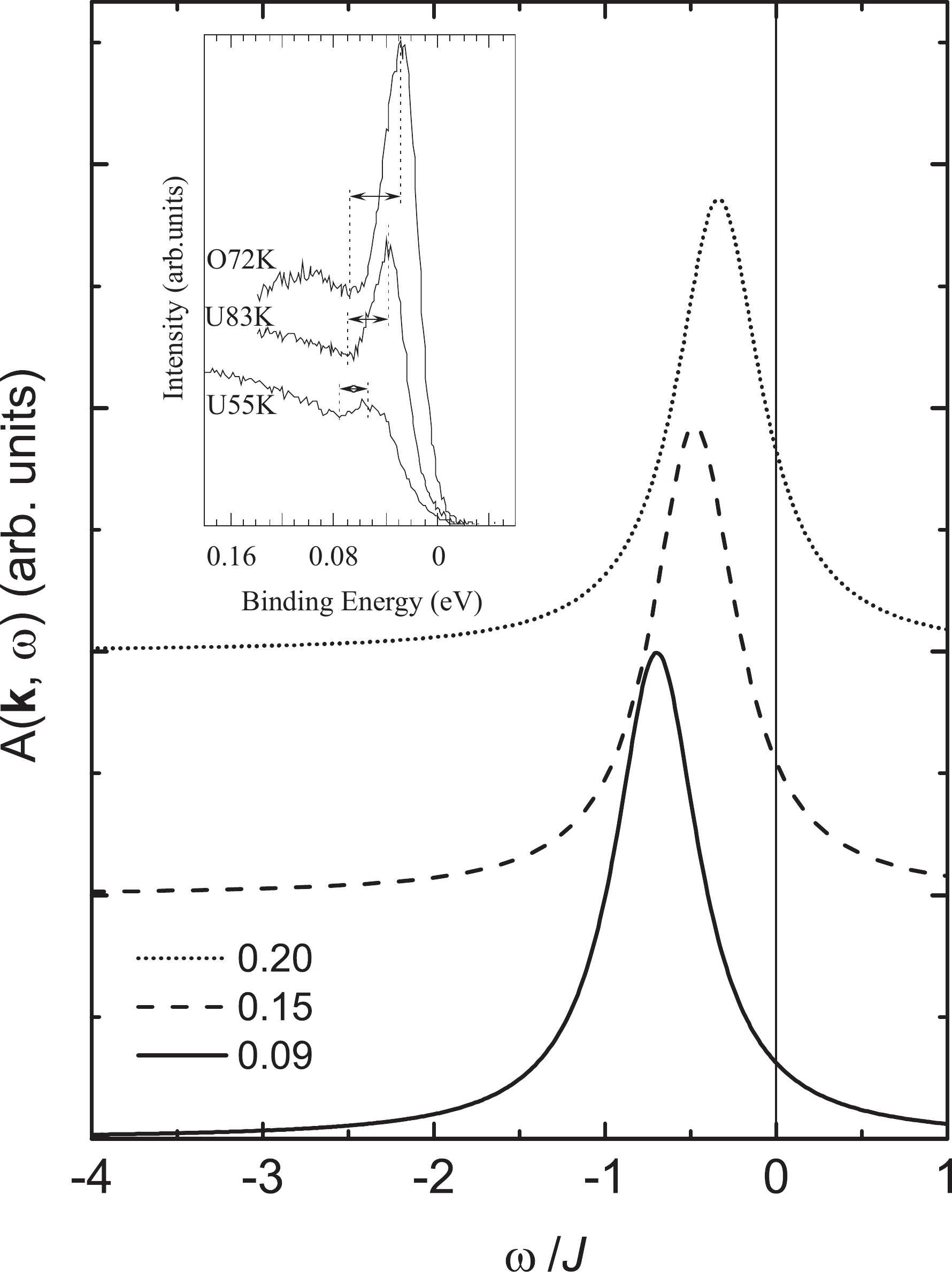}
\caption{The electron spectral function near the $[\pi,0]$ point at $\delta=0.09$ (solid line), $\delta=0.15$ (dashed line), and $\delta=0.20$ (dotted line) with $T=0.002J$ for $t/J=2.5$ and $t'/J=0.3$.  Inset: the corresponding experimental result of  Bi$_{2}$Sr$_{2}$CaCu$_{2}$O$_{8+x}$ taken from Ref. [\onlinecite{Campuzano99}]. \label{spectrum-doping}}
\end{figure}

The electron spectral function $A({\bf k},\omega)$ is obtained directly from the above electron diagonal Green's function (\ref{BCSEDGF}) as,
\begin{eqnarray}\label{spectral}
A({\bf k},\omega)&=&2\pi Z_{\rm F}[U^{2}_{\bf k}\delta(\omega-E_{\bf k})+V^{2}_{\bf k}\delta(\omega+E_{\bf k})],
\end{eqnarray}
where the height of the SC quasiparticle peak below the electron Fermi surface is assigned a weight $Z_{\rm F}V^{2}_{{\bf k}}$, while that of the peak above the electron Fermi surface is assigned a weight $Z_{\rm F}U^{2}_{{\bf k}}$, therefore the products of the electron coherence factors and the electron quasiparticle coherent weight describe the relative intensity of the Bogoliubov quasiparticle branches above and below the electron Fermi surface.

We are now in a position to compare the theoretical results derived based on the kinetic energy driven SC mechanism with existing ARPES data. We have performed a series of calculations for the electron spectral function $A({\bf k},\omega)$ in Eq. (\ref{spectral}) with different doping concentrations, and the result of $A({\bf k},\omega)$ near the antinodal $[\pi,0]$ point of the Brillouin zone (BZ) at the doping concentrations $\delta=0.09$ (solid line), $\delta=0.15$ (dashed line), and $\delta=0.20$ (dotted line) with temperature $T=0.002J$ for parameters $t/J=2.5$ and $t'/J=0.3$ is plotted in Fig. \ref{spectrum-doping} in comparison with the corresponding experimental result \cite{Campuzano99} of Bi$_{2}$Sr$_{2}$CaCu$_{2}$O$_{8+x}$ (inset). Our result shows that there are sharp low-energy SC quasiparticle peaks below the electron Fermi energy near the $[\pi,0]$ point, which are closely associated with electron Cooper pairing below $T_{\rm c}$. Using an reasonably estimative value of $J\sim 100$ meV in cuprate superconductors, the position of the low-energy SC quasiparticle peak at $\delta=0.15$ is located at $\omega_{{\rm peak}}\approx 0.5J\approx 50$ meV, which is not too far from the $\omega_{{\rm peak} }\approx 40$ meV observed \cite{Campuzano99} in the optimally doped Bi$_{2}$Sr$_{2}$CaCu$_{2}$O$_{8+x}$. However, the electron spectrum is doping dependent, and the weight of the low-energy SC quasiparticle peak in the underdoped regime increases with the increase of doping, while the position of the low-energy SC quasiparticle peak shifts towards to the electron Fermi energy, in qualitative agreement with the experimental results  \cite{Damascelli03,Campuzano04,Campuzano96,Matsui03,Balatsky07,Fujita08,Balatsky09,Campuzano99,Fedorov99,Ding01,DLFeng00}.

\begin{figure}[h!]
\centering
\includegraphics[scale=0.35]{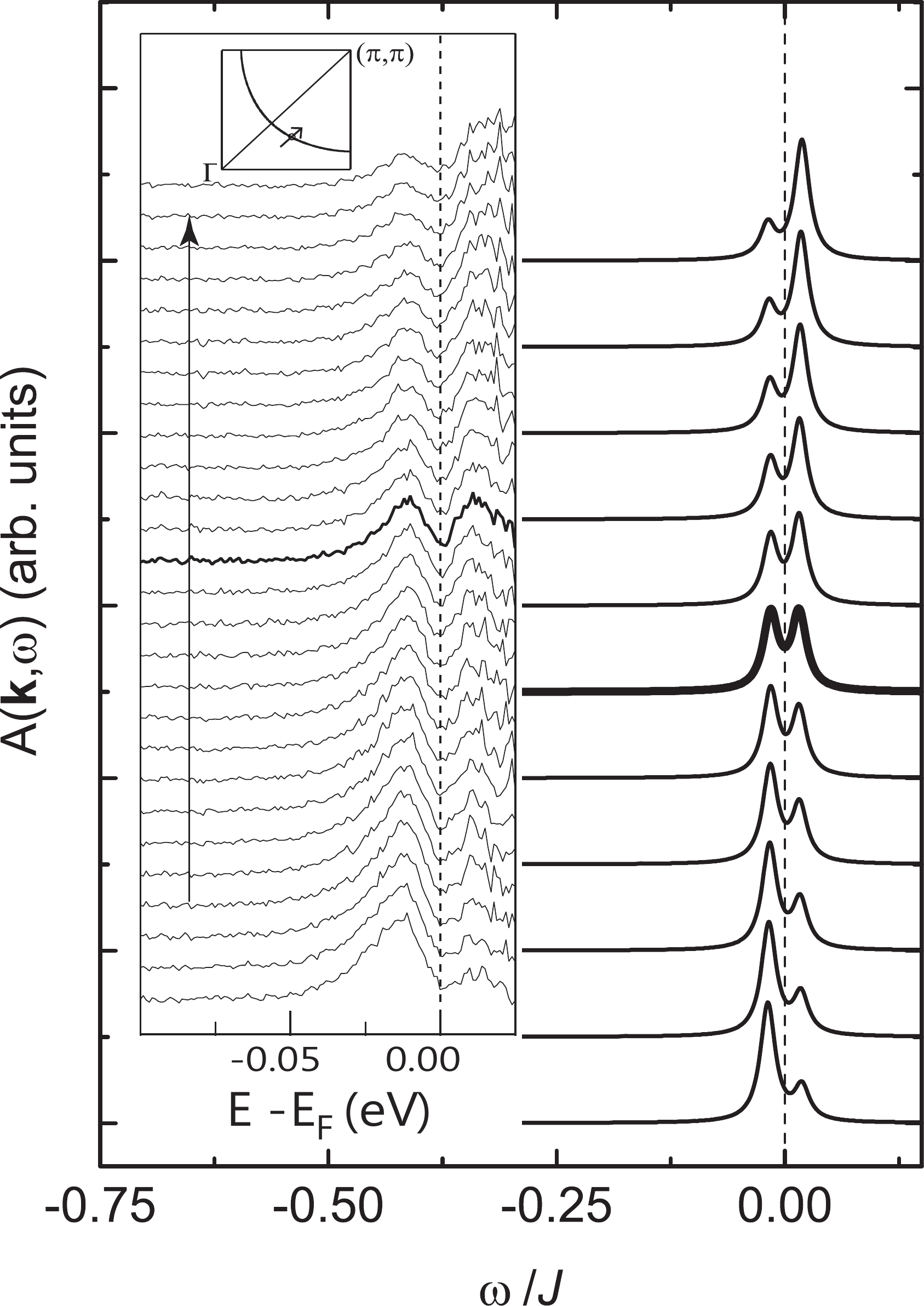}
\caption{The electron spectral function along the cut position $[0.679\pi,0.469\pi]$ to $[0.681\pi,0.471\pi]$ at $\delta=0.15$ with $T=0.002J$ for $t/J=2.5$ and $t'/t=0.3$. Inset: the corresponding experimental result of Bi$_{2}$Sr$_{2}$CaCu$_{2}$O$_{8+x}$ taken from Ref. [\onlinecite{Balatsky09}]. \label{spectrum-momentum}}
\end{figure}

For a complement of the analysis of the nature of the low-energy SC quasiparticle excitations, we have made a calculation for $A({\bf k},\omega)$ along the cut direction $[0.679\pi,0.469\pi]$ to $[0.681\pi,0.471\pi]$ crossing the electron Fermi surface just as it has been done in the ARPES experiments \cite{Balatsky09}, and the result of $A({\bf k}, \omega)$ as a function of energy with $T=0.002J$ for $t/J=2.5$ and $t'/J=0.3$ at $\delta=0.15$ is plotted in Fig. \ref{spectrum-momentum} in comparison with the corresponding experimental result \cite{Balatsky09} of the optimally doped Bi$_{2}$Sr$_{2}$CaCu$_{2}$O$_{8+\delta}$ (inset). In Fig. \ref{spectrum-momentum}, the thick solid curve is the momentum distribution curve where the electron coherence factors $U^{2}_{{\bf k}}=V^{2}_{{\bf k}}$ at the electron Fermi energy. It is apparent that the theoretical result captures the qualitative feature of the momentum dependence of the electron spectrum observed experimentally on cuprate superconductors in the SC-state \cite{Damascelli03,Campuzano04,Campuzano96,Matsui03,Balatsky07,Fujita08,Balatsky09,Campuzano99,Fedorov99,Ding01,DLFeng00}. There are two branches of dispersion centered at the electron Fermi energy, however, two sharp low-energy SC quasiparticle peaks in each energy distribution curve exhibit an evolution of the relative peak height at different momentum positions due to the momentum dependence of the coherence factors. In particular, the spectral intensity of the two branches shows an opposite evolution as a function of ${\bf k}$ along the cut direction $[0.679\pi,0.469\pi]$ to $[0.681\pi,0.471\pi]$. Before reaching the Fermi wave vector $k_{\rm F}$ along the  cut direction $[0.679\pi,0.469\pi]$ to $[0.681\pi,0.471\pi]$, the low-energy SC quasiparticle peak below the electron Fermi energy has a higher intensity than that above the electron Fermi energy. However, after passing the Fermi wave vector $k_{\rm F}$, the low-energy SC quasiparticle peak above the electron Fermi energy has a higher intensity than that below the electron Fermi energy. This crossover behavior near the Fermi wave vector $k_{\rm F}$, a characteristic of the Bogoliubov quasiparticle dispersion in conventional superconductors in the SC-state, therefore appears in cuprate superconductors.

\begin{figure}[h!]
\centering
\includegraphics[scale=0.35]{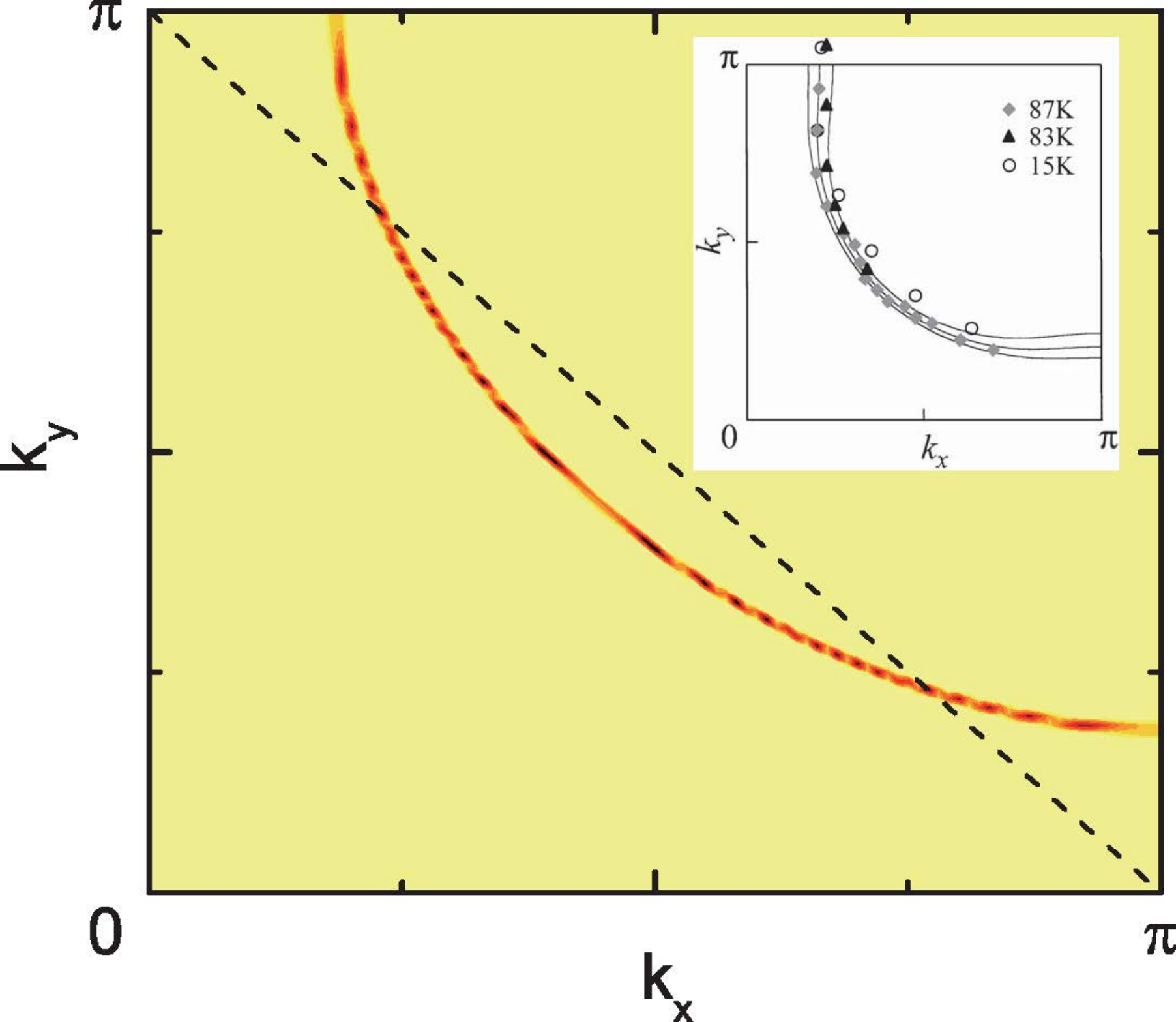}
\caption{(Color online) The spectral intensity maps at the electron Fermi energy in $\delta=0.12$ with $T=0.002J$ for $t/J=2.5$ and $t'/t=0.3$. Inset: the corresponding experimental data of Bi$_{2}$Sr$_{2}$CaCu$_{2}$O$_{8+x}$ taken from Ref. [\onlinecite{Ding97}]. \label{spectral-maps}}
\end{figure}

The notion of the Fermi surface is one of the characteristic concepts in the field of condensed matter physics, and it plays a crucial role in the understanding of the physical properties of interacting electron systems. This is why a central question in the theory of cuprate superconductors concerns the nature of the electron Fermi surface. In Fig. \ref{spectral-maps}, we plot the maps of the spectral intensity $A({\bf k},0)$ at the electron Fermi energy in $\delta=0.12$ for $t/J=2.5$ and $t'/t=0.3$ with $T=0.002J$ in comparison with the corresponding experimental result \cite{Ding97} of the underdoped Bi$_{2}$Sr$_{2}$CaCu$_{2}$O$_{8+x}$ (inset). Obviously, the electron Fermi surface forms a continuous contour in momentum space. In particular, we find that within the framework of the standard d-wave BCS formalism in Eqs. (\ref{BCSEGF}) and (\ref{BCSECF}), the electron Fermi surface in the entire doping range forms a continuous contour in momentum space \cite{Takahashi89,Campuzano90,Olson90,Ding97}. Moreover, according to one of the self-consistent equations (\ref{ESCE2}), the electron Fermi surface satisfies Luttinger's theorem, i.e., the electron Fermi surface area contains $1-\delta$ electrons. However, we will show in the following discussions that in the underdoped and overdoped regimes, the energy and momentum dependence of the electron self-energy in the particle-hole channel truncates this continuous contour in momentum space into the disconnected Fermi arcs located around the nodal region of BZ.

All above theoretical results are qualitatively consistent with the ARPES experimental data of cuprate superconductors \cite{Campuzano96,Matsui03,Balatsky07,Fujita08,Balatsky09,Campuzano99,Fedorov99,Ding01,DLFeng00,Takahashi89,Campuzano90,Olson90,Ding97}, and therefore show that the basic d-wave BCS formalism obtained from the kinetic energy driven SC mechanism can correctly reproduce some main features of the SC coherence of the low-energy quasiparticle excitations observed in cuprate superconductors in the SC-state.

\subsection{Effect of energy and momentum dependence of self-energy on electron spectrum}\label{self-energy-electronic-structure}

It is well known that the electron correlation in solids is closely related to the electron self-energy. In particular, the positions of the quasiparticle peaks in the electron spectrum are determined by the electron self-energy. Although the d-wave BCS expression in Eq. (\ref{spectral}) gives qualitative insight into the electronic spectrum in the SC-state, the detailed form of the line shape that arises from the strong energy and momentum dependence of the electron self-energy is not contained in the simple d-wave BCS formalism (\ref{BCSEGF}). In the following discussions, we show explicitly that the striking feature of the peak-dip-hump structure in the electron spectrum of cuprate superconductors and the remarkable Fermi arc behavior in the underdoped and optimally doped regimes are intimately connected with the strong energy and momentum dependence of the electron self-energy. In order to take into account the effect of the energy and momentum dependence of the electron self-energy on the electron spectrum in the SC-state, the full electron diagonal and off-diagonal Green's functions in Eq. (\ref{EGF}) can be rewritten as,
\begin{widetext}
\begin{subequations}\label{PGEGF}
\begin{eqnarray}
G({\bf k},\omega)&=&{1\over \omega-\varepsilon_{\bf k}-\Sigma_{1}({\bf k},\omega)-[\bar{\Delta}({\bf k})]^{2}/[\omega+\varepsilon_{\bf k}+\Sigma_{1}({\bf k},-\omega)]}, \label{PGEDGF}\\
\Im^{\dagger}({\bf k},\omega)&=&-{\bar{\Delta}({\bf k})\over [\omega-\varepsilon_{\bf k}-\Sigma_{1}({\bf k},\omega)][\omega+\varepsilon_{\bf k}+\Sigma_{1}({\bf k},-\omega)]
-[\bar{\Delta}({\bf k})]^{2}},
\end{eqnarray}
\end{subequations}
\end{widetext}
where the energy and momentum dependence of the electron self-energy $\Sigma_{1}({\bf k},\omega)$ in the particle-hole channel has been given explicitly in Eq. (\ref{PHESE}), and then the electron spectral function,
\begin{eqnarray}
A({\bf k},\omega)=-2{\rm Im}G({\bf k},\omega), \label{SE-spectral}
\end{eqnarray}
can be evaluated directly from the full electron diagonal Green's function (\ref{PGEDGF}).

\begin{figure}
\centering
\includegraphics[scale=0.35]{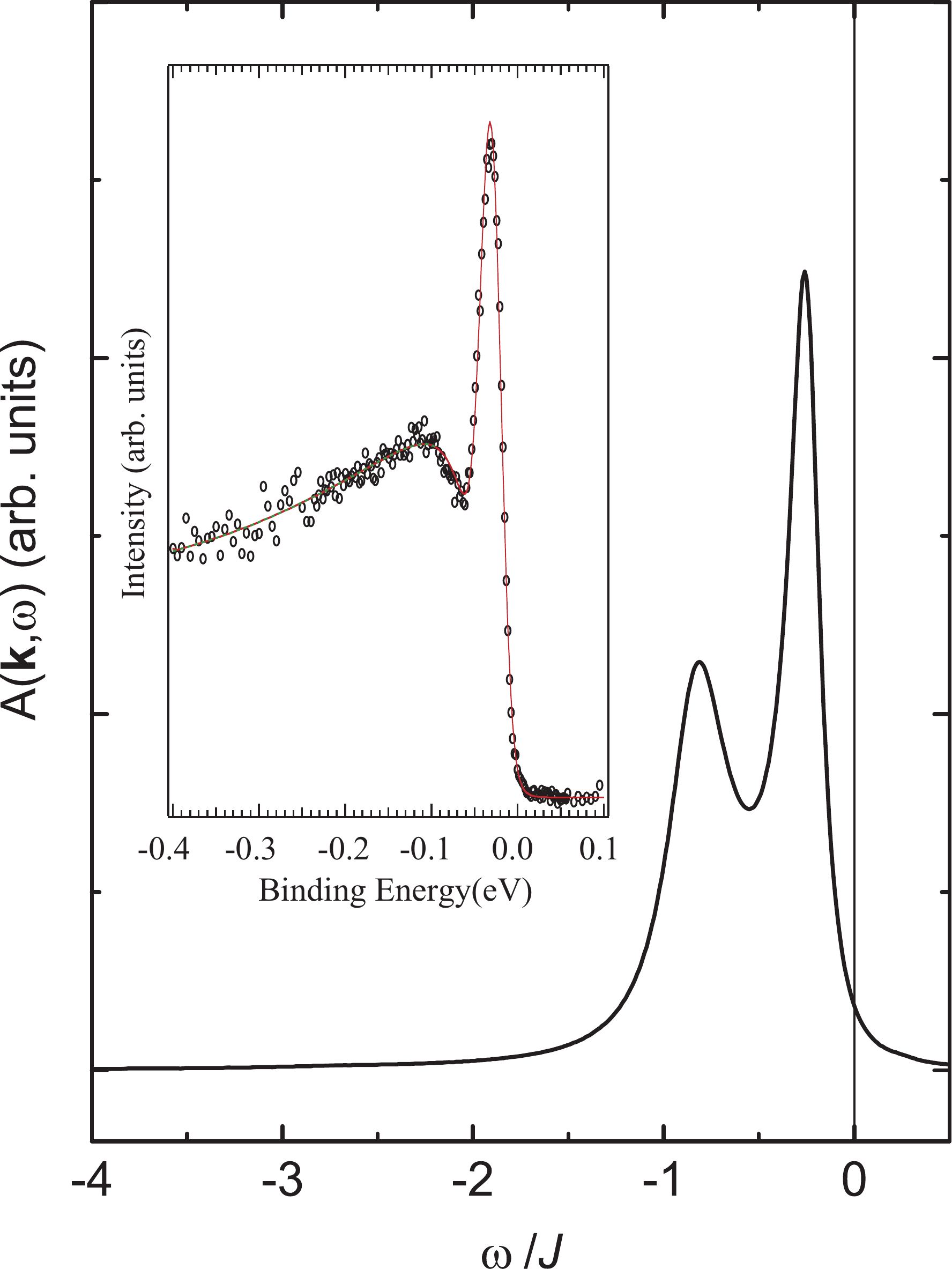}
\caption{The electron spectral function near the $[\pi,0]$ point at $\delta=0.20$ with $T=0.002J$ for $t/J=2.5$ and $t'/t=0.3$.  Inset: the corresponding experimental data of  Bi$_{2}$Sr$_{2}$CaCu$_{2}$O$_{8+x}$ taken from Ref. [\onlinecite{Ding01}]. \label{spectrum-PDH}}
\end{figure}

Although the sharp quasiparticle excitations are identified in the SC-state along the entire electron Fermi surface, the peak-dip-hump structure is most strongly developed around the $[\pi,0]$ point. In Fig. \ref{spectrum-PDH}, we plot the electron spectral function $A({\bf k},\omega)$ in Eq. (\ref{SE-spectral}) near the $[\pi,0]$ point for $t/J=2.5$ and $t'/t=0.3$ with $T=0.002J$ at $\delta=0.20$ in comparison with the corresponding experimental result \cite{Ding01} of the overdoped Bi$_{2}$Sr$_{2}$CaCu$_{2}$O$_{8+x}$ (inset), where the main feature of the electron spectral function in cuprate superconductors \cite{Campuzano99,Fedorov99,Ding01} is qualitatively reproduced. In particular, the position of the low-energy SC quasiparticle peak is located at $\omega_{{\rm peak}}\approx 0.32J\approx 32$ meV, which is well consistent with the $\omega_{{\rm peak}}\approx 35$ meV observed \cite{Ding01} in the overdoped Bi$_{2}$Sr$_{2}$CaCu$_{2}$O$_{8+x}$. In comparison with the corresponding result in Fig. \ref{spectrum-doping}, we therefore find that an additional peak in the electron spectrum appears at the higher energy region, however, the weight of this additional peak is much smaller than that in the lower-energy region. In this case, the electron spectrum consists of two peaks, with a sharp low-energy peak that is corresponding to the SC coherence of the quasiparticle excitation, and a weak high-energy peak, which is associated with the hump, while the spectral dip is in between them, and then the total contributions for the electron spectrum give rise to the peak-dip-hump structure, in good agreement with the ARPES experimental observations on cuprate superconductors \cite{Damascelli03,Campuzano04,Campuzano99,Fedorov99,Ding01,DLFeng00}. The present result in Fig. \ref{spectrum-PDH} also indicates that the existence of the dip requires additional structure in the electron self-energy, and then the striking feature of the peak-dip-hump structure in the single-particle excitation spectrum is closely related to the strong energy and momentum dependence of the electron self-energy. This is different from the case in conventional superconductors, where the electron self-energy has essentially no momentum dependence.

\begin{figure}[h!]
\centering
\includegraphics[scale=0.37]{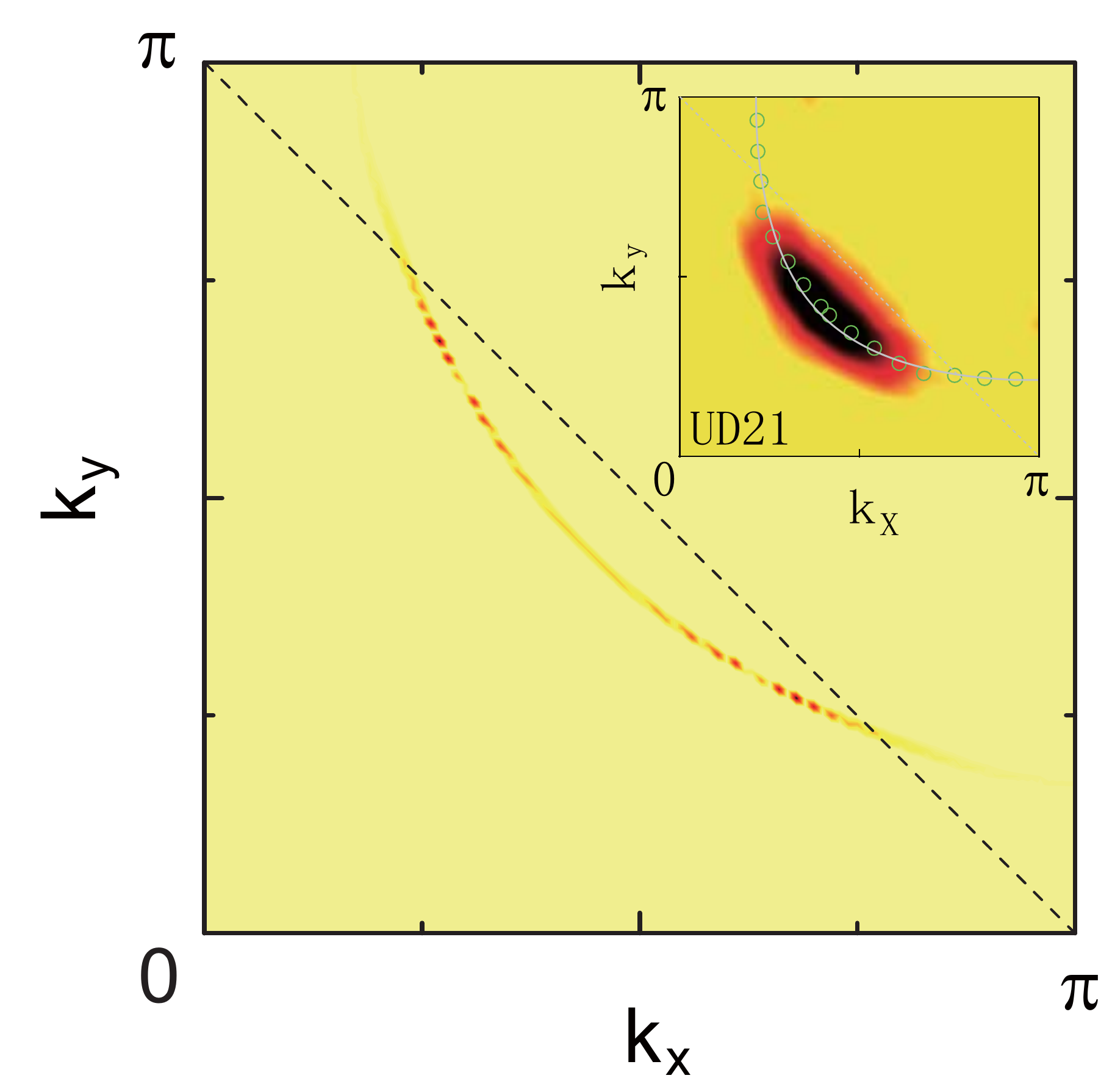}
\caption{(Color online) The spectral intensity maps at the electron Fermi energy in $\delta=0.12$ for $t/J=2.5$ and $t'/t=0.3$ with $T=0.002J$. Inset: the corresponding experimental data of Ca$_{2-x}$Na$_{x}$CuO$_{2}$Cl$_{2}$ taken from Ref. [\onlinecite{Meng11}]. \label{SE-spectral-maps}}
\end{figure}

To analyze the effect of the strong energy and momentum dependence of the electron self-energy on the electron Fermi surface, we plot the electron spectral density $A({\bf k},0)$ in Eq. (\ref{SE-spectral}) at the Fermi energy in $\delta=0.12$ for $t/J=2.5$ and $t'/t=0.3$ with $T=0.002J$ in Fig. \ref{SE-spectral-maps}. For comparison, the corresponding experimental result \cite{Meng11} of the underdoped Ca$_{2-x}$Na$_{x}$CuO$_{2}$Cl$_{2}$ is also shown in Fig. \ref{SE-spectral-maps} (inset). In comparison with the corresponding result in Fig. \ref{spectral-maps}, the present result in Fig. \ref{SE-spectral-maps} shows clearly that the electron Fermi surface in the underdoped regime does not form a continuous contour in momentum space due to the presence of the energy and momentum dependence of the electron self-energy, where the electron self-energy suppression first opens up near the antinodal region, and then the electron Fermi surface is broken into the disconnected Fermi arc around the nodal region. In particular, we \cite{Kuang15a} have made a series of calculations for the electron spectral density $A({\bf k},0)$ at the Fermi energy with different doping concentrations, and the result shows that the strength of the electron self-energy suppression progressively decreases as the doping concentration is increased, and then there is a tendency towards to form a continuous contour in momentum space. This tendency is particularly obvious in the overdoped regime, and then the electron Fermi surface evolves into a continuous contour in momentum space at the heavily overdoped regime, in qualitative agreement with experimental data  \cite{Norman98,Kanigel06,Nakayama09,Meng11,Yoshida06}.

\begin{figure}[h!]
\centering
\includegraphics[scale=0.37]{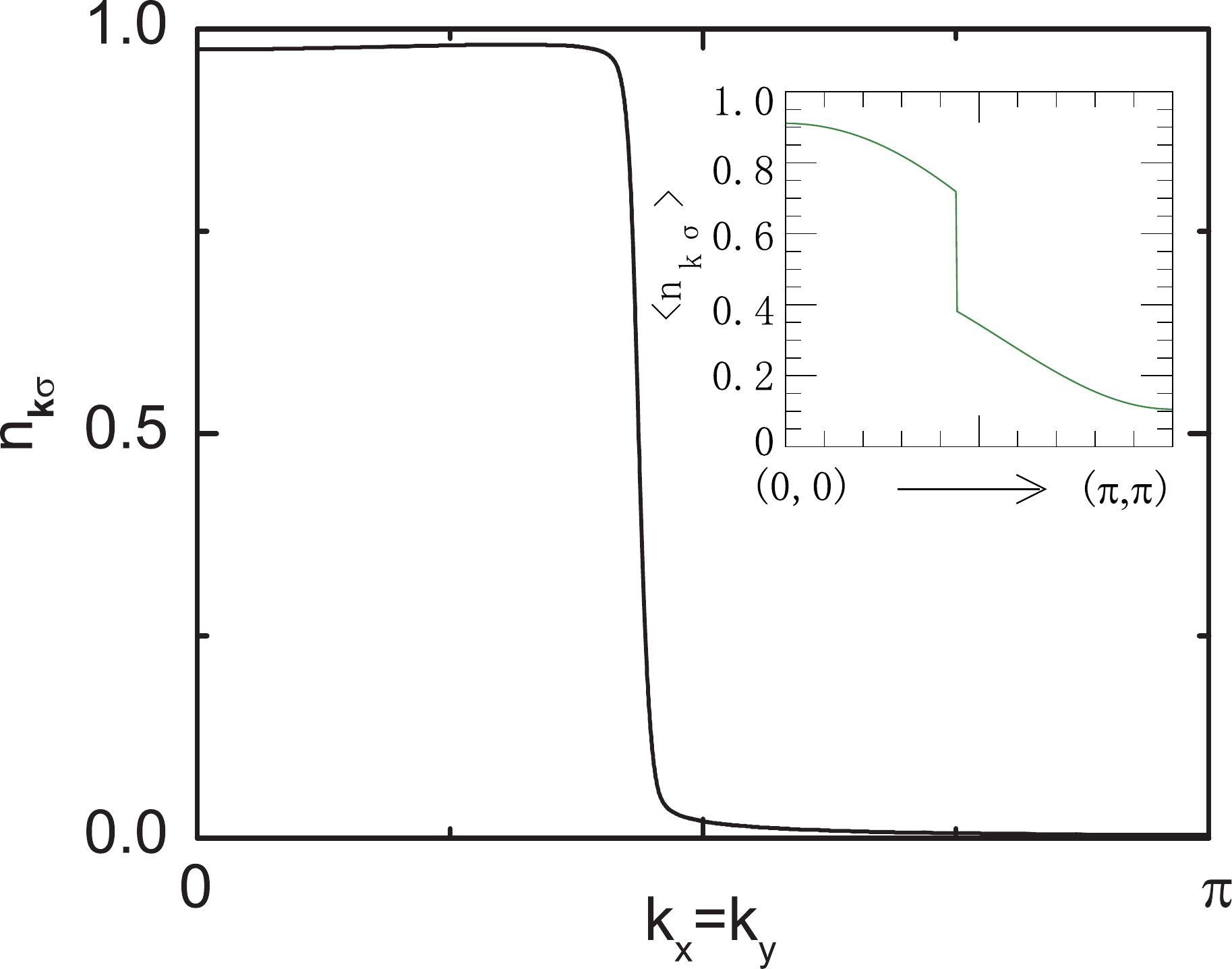}
\caption{The electron-momentum distribution along the ${\bf k}=[0,0]$ to ${\bf k}=[\pi,\pi]$ direction at $\delta=0.12$ with $T=0.002J$ for $t/J=2.5$ and $t'/t=0.3$. Insets: the corresponding numerical result taken from Ref. [\onlinecite{Edegger06}] \label{SE-momentum-distribution}}
\end{figure}

Now we turn to discuss the electron momentum distribution. The number of electrons in state {\bf k} is obtained by summing over all energies $\omega$, weight by the electron spectral function (\ref{SE-spectral}) as,
\begin{eqnarray}
n_{{\bf k}\sigma}= \int^{\infty}_{-\infty}{{\rm d}\omega\over 2\pi}n_{\rm F}(\omega)A({\bf k},\omega) \label{SE-number}.
\end{eqnarray}
For a better understanding of the nature of the electron Fermi surface in the kinetic energy driven SC mechanism, we plot the electron momentum distribution $n_{{\bf k}\sigma}$ in Eq. (\ref{SE-number}) along the ${\bf k} = [0,0]$ to ${\bf k}= [\pi,\pi]$ direction at $\delta=0.12$ with $T=0.002J$ for $t/J=2.5$ and $t'/t=0.3$ in Fig. \ref{SE-momentum-distribution} in comparison with the corresponding numerical result \cite{Edegger06} obtained from a variational Monte Carlo technique. Since the total number of electrons does not depend on the interaction, the integrated area under the curve is equal to $1-\delta$, and therefore is consistent with that predicted by the Luttinger's theorem. The shape of the electron momentum distribution on the other hand is a should-be electron momentum distribution of the system in the presence of interactions, i.e., in some part (below the electron Fermi energy) the distribution is closer to $1$, while in other part (above the electron Fermi energy) it is approximately closer to zero, in qualitative agreement with the results obtained from the numerical simulations \cite{Paramekanti01,Edegger06,Stephan91}.

The above calculated result based on the kinetic-energy driven SC mechanism is able to give qualitative description of the electronic structure of cuprate superconductors in the SC-state. In this case, a natural question is: What is the reason why the electronic structure of cuprate superconductors in the SC-state can be described qualitatively within the framework of the kinetic-energy driven superconductivity. To our present understanding, there are, at least, two reasons: (A) Although the physical properties of cuprate superconductors in the normal-state are fundamentally different from these in the standard Landau Fermi-liquid state, the kinetic-energy driven SC-state still is conventional BCS-like with the d-wave symmetry; (B) The emergence of the doping dependence of the pseudogap is essential. In our previous work \cite{Feng12}, we have shown that the same charge-carrier interaction mediated by spin excitations that induces the charge-carrier pairing state in the particle-particle channel also generates the pseudogap state in the particle-hole channel, and therefore the pseudogap has been identified as being a region of the charge-carrier self-energy in the particle-hole channel in which the pseudogap suppresses the low-energy spectral weight of the quasiparticle excitation spectrum. Since the electron Cooper pairing state originated from the charge-carrier pairing state is due to the full charge-spin recombination, the physics of the pseudogap state in the charge-carrier pairing state is also true in the present electron Cooper pairing state. This follows from a fact that the electron self-energy $\Sigma_{1}({\bf k},\omega)$ in the particle-hole channel in Eq. (\ref{PHESE}) also can be rewritten approximately as,
\begin{eqnarray}\label{EPG}
\Sigma_{1}({\bf k},\omega)&\approx&{[\bar{\Delta}_{\rm PG}({\bf k})]^{2}\over \omega+\varepsilon_{0{\bf k}}},
\end{eqnarray}
where $\varepsilon_{0{\bf k}}=L^{({\rm e})}_{2}({\bf k})/L^{({\rm e})}_{1}({\bf k})$ is the energy spectrum of $\Sigma_{1}({\bf k},\omega)$, and $\bar{\Delta}_{\rm PG} ({\bf k})= L^{({\rm e})}_{2}({\bf k})/[2\sqrt{L^{({\rm e})}_{1}({\bf k})}]$ is the pseudogap, with the functions $L^{({\rm e})}_{1}({\bf k})=-\Sigma_{\rm 1o}({\bf k},\omega=0)$ and $L^{({\rm e})}_{2} ({\bf k})=\Sigma_{1}({\bf k},\omega=0)$, and can be obtained directly from the electron self-energy $\Sigma_{1}({\bf k},\omega)$ in Eq. (\ref{PHESE}). This pseudogap $\bar{\Delta}_{\rm PG}({\bf k})$ is also identified as being a region of the electron self-energy in the particle-hole channel in which the pseudogap $\bar{\Delta}_{\rm PG}({\bf k})$ suppresses the low-energy spectral weight of the single-particle excitation spectrum. This pseudogap (then the electron self-energy) therefore induces the appearance of the additional high-energy peak in the electron spectrum, leading to the formation of the peak-dip-hump structure, and gaps out the antinodal region of the electron Fermi surface, leading to that the electron Fermi surface consists, not of closed contour, but only of four disconnected Fermi arcs centered around the nodes. In particular, we \cite{Kuang15a} find that the present pseudogap $\bar{\Delta}_{\rm PG}({\bf k})$ in Eq. (\ref{EPG}) has the same doping dependence as that obtained previously from the charge-carrier self-energy in Ref. \cite{Feng12}, i.e., the magnitude of the pseudogap parameter $\bar{\Delta}_{\rm PG}$ is particular large in the underdoped regime, and then it smoothly decreases upon increasing doping, in qualitative agreement with the ARPES experimental results \cite{Yoshida09}. This doping dependence of the pseudogap $\bar{\Delta}_{\rm PG}({\bf k})$ therefore leads to an evolution of the Fermi arcs with doping in cuprate superconductors \cite{Norman98,Kanigel06,Nakayama09,Meng11,Yoshida06}.

\section{Conclusions}\label{conclusions}

Within the framework of the kinetic energy driven SC mechanism, we have developed a full charge-spin recombination scheme. Following this full charge-spin recombination scheme, we have evaluated explicitly the full electron diagonal and off-diagonal Green's functions, and then reproduced some main features of the electronic structure in cuprate superconductors in the SC-state. Our result shows that the theory produces a large electron Fermi surface with the area that contains $1-\delta$ electrons, while the SC coherence of the low-energy quasiparticle excitations is qualitatively described by the standard d-wave BCS formalism, although the pairing mechanism is driven by the kinetic energy by the exchange of spin excitations in the higher powers of the doping concentration. Our result also shows that the striking peak-dip-hump structure in the electron spectrum and the remarkable Fermi arc behavior in the underdoped regime are mainly caused by the strong energy and momentum dependence of the electron self-energy. The qualitative agreement between the theoretical result and ARPES experimental data is important to confirm that the SC-state of cuprate superconductors still is conventional BCS-like with the d-wave symmetry.

\acknowledgments

The authors would like to thank Yu Lan, Ling Qin, and Xixiao Ma for helpful discussions. This work was supported by the funds from the Ministry of Science and Technology of China under Grant Nos. 2011CB921700 and 2012CB821403, and the National Natural Science Foundation of China under Grant Nos. 11274044 and 11447144.

\end{document}